\documentclass[a4paper]{mem}
\usepackage{natbib}
\usepackage{graphicx}
\usepackage[a4paper]{hyperref}

\begin{document}


\title{An IR view of mass loss in Long Period Variables}

\author{M. Marengo \inst{1}, M. Karovska \inst{1},
        \and \v Z. Ivezi\'c \inst{2}}

\offprints{M. Marengo}
\mail{M. Marengo, email address mmarengo@cfa.harvard.edu }

\institute{Harvard-Smithsonian Center for Astrophysics, 
           Cambridge, MA (USA)
      \and Princeton University, Dep. of Astrophysical Sciences.
           Princeton, NJ (USA)
          }

\authorrunning{Marengo et al.}
\titlerunning{Mass loss in LPV}

\abstract{We have investigated the mass loss processes in Long Period
Variables, by combining radiative transfer modeling with mid-IR
imaging and spectroscopic observations. We find a correlation between
variability type and temporal variations in the mass loss rate. Mira
variables are more likely to maintain constant mass loss rates over
long period of time. Semiregular and Irregular variables are
characterized by frequent interruptions in their mass loss rate, on
timescales of $\sim 100$~yr. High resolution imaging of individual
Long Period Variable sources show departures from spherical
symmetry. We present the case of $o$~Cet (Mira), in which the presence
of a low mass companion plays an active role in shaping the
circumstellar environment.

\keywords{circumstellar matter -- stars: mass-loss - stars: variables:
other -- infrared: stars -- stars: individual ($o$ Ceti)}

}

\maketitle


\section{Introduction}

Intense mass loss is an important factor in the late evolutionary
phases of Long Period Variables (LPVs). Mass loss processes are
responsible for the creation of circumstellar envelopes of gas and
dust around these stars, which will later evolve into Planetary
Nebul\ae{} (PNs). HST images of PNs reveal a rich variety of bipolar
structures, including jets, shell fragments and asymmetries, which are
largely unexpected in objects evolved from spherically symmetric
stars. Understanding the origin of such asymmetries is thus of a crucial
interest for the comprehension of mass loss processes in the latest
stages of stellar evolution, and specifically in the precursor stars
of PNs. This analysis is best done at infrared wavelengths where dust
thermal emission peaks, since the dynamics of LPV circumstellar
envelopes is dominated by the dust grains (``dust driven winds'').

With this in mind, we have started a long term study of LPVs,
combining radiative transfer modeling with observations performed in
the mid-IR. Our results show that temporal variations in the mass loss
rate and deviations from spherical symmetry are in fact common in PN
precursors. 

\begin{figure*}[t]
\centering
\includegraphics[width=0.9\textwidth]{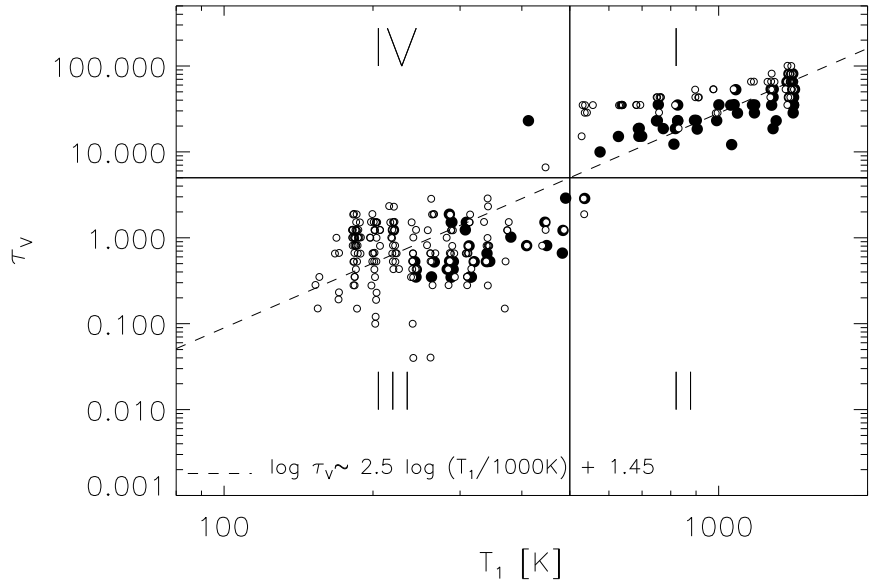}
\caption{\footnotesize{Best fit parameters for our sources: Miras (filled
circles) and non-Miras (open circles).}}\label{fig1}
\end{figure*}


\section{Radiative transfer modeling}

The spectral energy distribution and broad band colors at mid-IR
wavelengths provide a reliable diagnostic tool for the spatial
distribution and thermodynamical conditions of LPV circumstellar dust
\citep[see e.g.][]{marengo1999}. The two main parameters that can be
derived with such analysis are the total optical depth $\tau_V$ and
the temperature structure of the dust circumstellar envelope. 
The temperature $T_1$ at the inner edge of the
envelope, in particular, is an indicator of either the dust
condensation temperature $T_{cond}$, or the presence of an inner
cavity depleted of hot dust. Since central cavities in circumstellar
envelopes are generally associated with sudden interruption in the
mass loss processes, the determination of $T_1$ for a large sample of
LPVs would provide a method to study the temporal variations
of the mass loss rate in the sample.

For these reasons we have selected a sample of 342 LPVs of which
96 Mira, 188 Semiregular and 58 Irregular variables. The sources were
derived from a sample compiled by \citet{kerschbaum1996} with
additional requirements including the availability of the IRAS
Low Resolution Spectra (LRS) and O-rich chemistry. The last
requirement was enforced in order to have a homogeneous sample, 
and because of the better diagnostics provided by the
prominent 10~$\mu$m silicate feature. Following \citet{ivezic1997},
we have modeled the shape of the dust emission
features for all sources, deriving an estimate for $\tau_V$
and $T_1$. The details of our technique and a complete statistical
analysis of the results are presented in \citet[~hereafter
MIK]{marengo2001a}\defcitealias{marengo2001a}{MIK}.

Figure~\ref{fig1} shows the main result of our analysis: there is a
dichotomy between Mira and non-Mira variables, with 70\% of the Miras
having high optical depth and high dust temperature, and 70\% of the
non-Miras having low $\tau_V$ and temperature much below $T_{cond}$.
This result can be interpreted as an evidence that Semiregulars and
Irregulars are subjected to phases of reduced mass loss, in which the
envelope detaches and expands at the stellar wind velocity. This
hypothesis is supported by the best fit of the sources distribution,
which is consistent with the theoretical prediction $\tau_V \sim
T_1^{2.5}$ \citepalias{marengo2001a}. 

\begin{figure}[t]
\centering
\includegraphics[width=0.5\textwidth]{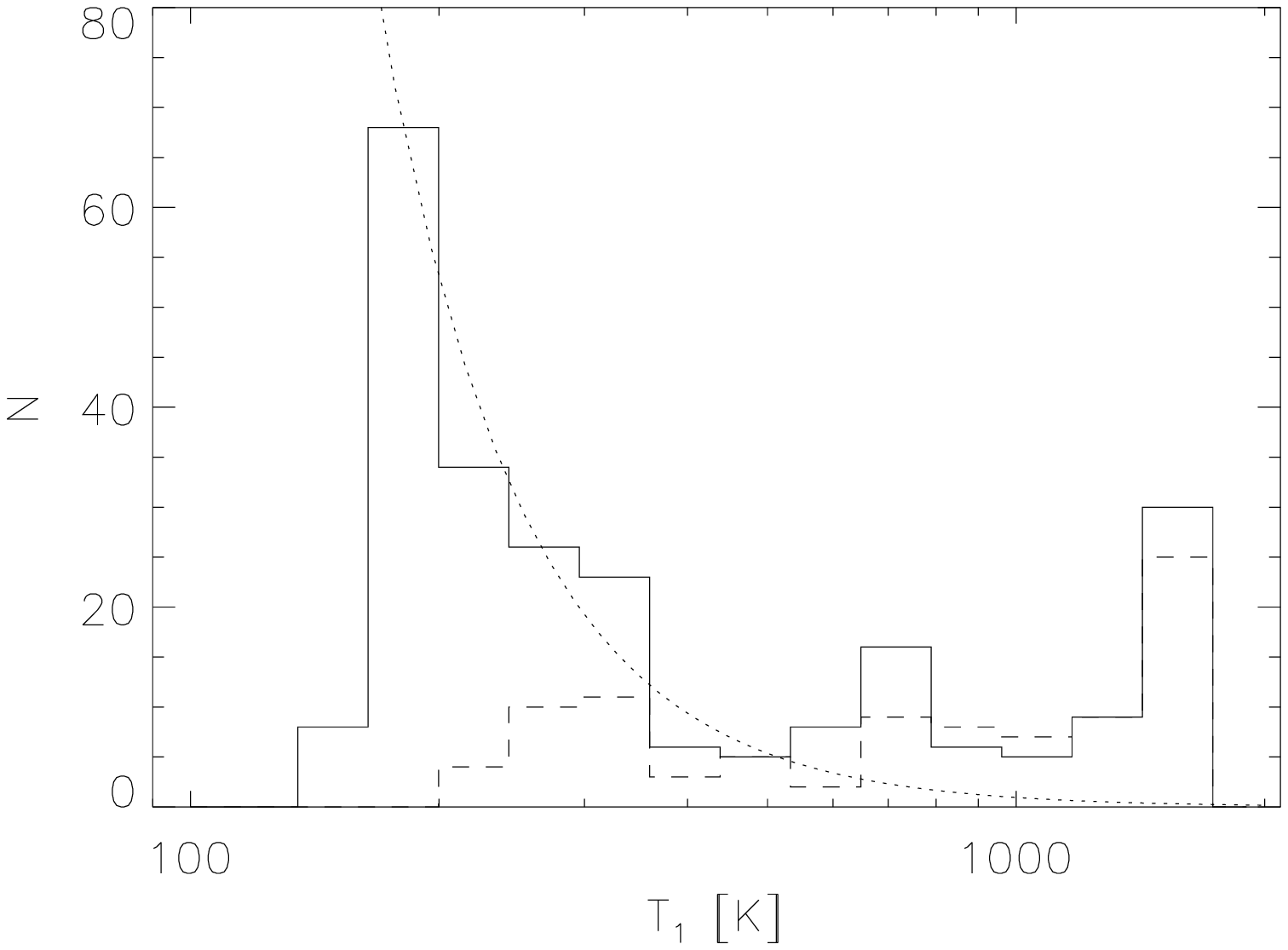}
\includegraphics[width=0.5\textwidth]{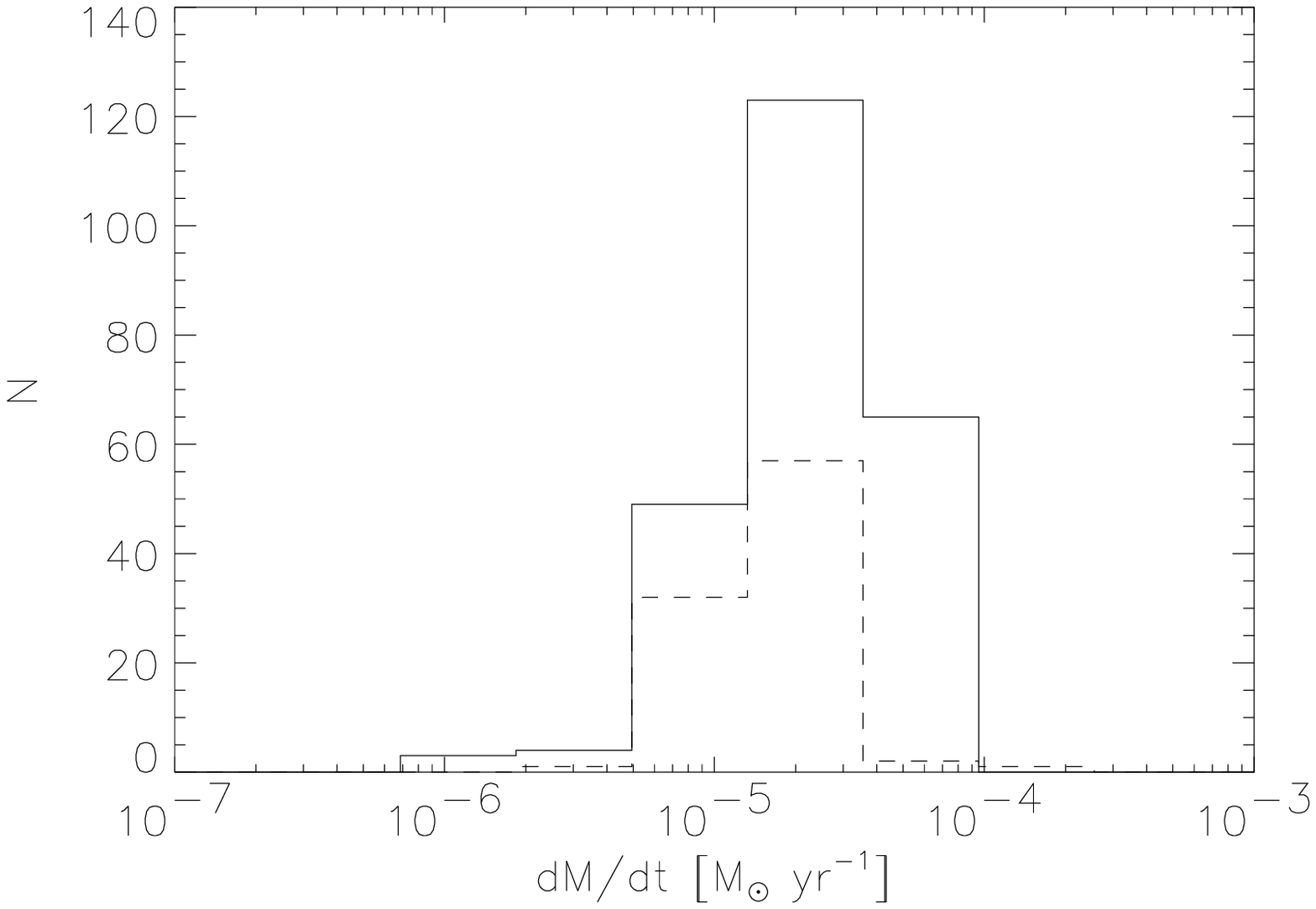}
\caption{\footnotesize{ Distribution of best fit $T_1$ and mass loss
rates for our sources: Miras (dashed line) and non-Miras (solid
line). The dotted line is the theoretical distribution for detached
shell.}}\label{fig2}
\end{figure}

The two histograms in Figure~\ref{fig2} show the distribution of $T_1$
and the mass loss rate $d M / d t$ for all sources. Miras show a
flattened distribution at the dust condensation temperature range. The
distribution of non-Miras, instead, follows a $dN/dt \sim T_1^{-3.5}$
law, as expected for detached envelopes \citepalias{marengo2001a}. The
minimum temperature of $\sim 150$~K for the coldest envelopes
suggests a timescale of $\sim 100$~yr as maximum duration of the
reduced mass loss phase. The histogram of the source mass loss rate,
on the contrary, does not show any difference between the two classes,
suggesting a similar mass loss during the active phase of dust
production.

In summary, this analysis suggests a different behavior in the mass
loss of Mira and non-Mira type LPVs. Mira circumstellar envelopes
are characterized by hotter dust which is an evidence of active dust
production and mass loss. In many Semiregular and Irregular variables,
on the contrary, cold dust prevails, and there are evidences of
detached shells as a consequence of a temporal reduction in the
mass loss rate, on timescales of $\sim 100$~yr. These differences does
not seem to be related to different mechanisms of dust production, 
but rather to the specific ability of maintaining a constant mass loss
rate of the two classes of pulsators. Confirmation of
this hypothesis can only be provided by a comprehensive theory of 
dust driven mass loss in LPVs, including dust condensation and
two-fluid time-dependent hydrodynamics. Important steps in this
directions have been made by J.M. Winters (private comm.) and
\citet{simis2001}.


\begin{figure*}[t]
\centering
\includegraphics[height=1\textwidth,angle=-90]{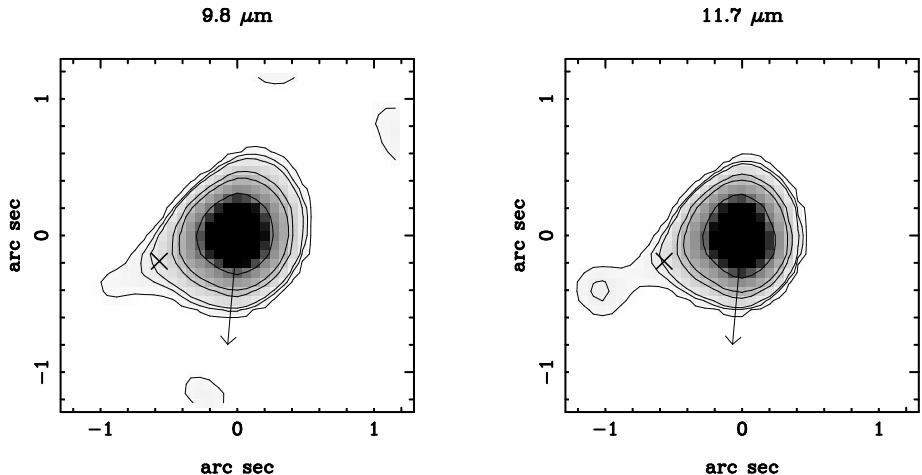}
\caption{\footnotesize{9.8~$\mu$m (left) and 11.7~$\mu$m (right)
images of the Mira system. The arrow indicate the main asymmetry of
Mira HST TiO images, and the cross shows the position of the
companion.}}\label{fig3}
\end{figure*}

\section{Mid-IR imaging}

High resolution imaging can in principle provide a direct test for the
existence of departures from ``steady-state'' and spherically symmetric
mass loss in LPVs. Such structures can be investigated in the mid-IR
by direct imaging of the thermal emission from circumstellar dust.
For this reason we are carrying-on a
long term campaign of mid-IR imaging of circumstellar envelopes around
nearby LPVs. The project started with the cameras TIRCAM and CAMIRAS
at the TIRGO (Switzerland) and San Pedro Martir (Mexico)
observatories, and is continuing with the MIRAC camera at the IRTF
(Mauna Kea), MMT (Arizona) and Magellan (Chile) telescopes. 

Preliminary results of our observations have been published in
\citet{marengo2000}. Of the MIRAC sources alone, 65\% of the total
show extended emission with total flux 20\% above the level of a point
source. Many sources have circumstellar envelopes with sizes as large
as 4~arcsec, departing from spherical symmetry. In one case (W~Hya) we
have detected evidences of a detached envelope. In another ($o$~Cet,
Mira AB) we have documented the case of a binary system in which the
companion star is playing an active role in shaping the common
circumbinary environment. 

Figure~\ref{fig3} shows the MIRAC/IRTF images of $o$~Cet at 9.8 and
11.7~$\mu$m. The source appears dominated by an asymmetry in the N-S
direction (PA $\sim 175$ deg), which is consistent with the main
asymmetry observed in the Mira A TiO envelope by HST
\citep{karovska1997}. The dusty envelope appears also extended in the
direction of Mira B, with a separate clump at PA $\sim 110$ deg,
suggesting that Mira A envelope may be shaped by the interaction with
the companion. These observations are described in detail in
\citet{marengo2001b}.


\section{Conclusions}

Our analysis show that temporal changes in the mass loss rate of
LPVs may be common, and correlated with the variability type.
Mira variables show the ability to maintain constant mass loss, while
non-Miras show the tendency to develop detached shells.
Whenever the angular resolution is sufficient, we have been able
to find indication that the dust formation in LPVs extended
atmospheres produces departures from spherical symmetry. In some cases
this asymmetries may be enhanced, or even determined, by the presence
of a companion, as in the $o$~Cet system.


\begin{acknowledgements}
M.K. is member of the Chandra Science Center, which is operated under
contract NAS8-39073 and is partially supported by NASA.
\end{acknowledgements}


\bibliographystyle{aa}

\end{document}